# DOLLmC: DevOps for Large Language model Customization

**Panos Fitsilis**[*], Business Administration Department, University of Thessaly, Greece

**Vyron Damasiotis**, Department of Accounting and Finance, University of Thessaly, Greece

**Vasileios Kyriatzis**, Department of Digital Systems, University of Thessaly, Greece

**Paraskevi Tsoutsa**, Department of Accounting and Finance, University of Thessaly, Greece

## Abstract

The rapid integration of Large Language Models (LLMs) into various industries presents both revolutionary opportunities and unique challenges. This research aims to establish a scalable and efficient framework for LLM customization, exploring how DevOps practices should be adapted to meet the specific demands of LLM customization. By integrating ontologies, knowledge maps, and prompt engineering into the DevOps pipeline, we propose a robust framework that enhances continuous learning, seamless deployment, and rigorous version control of LLMs. This methodology is demonstrated through the development of a domain-specific chatbot for the agricultural sector, utilizing heterogeneous data to deliver actionable insights. The proposed methodology, so called DOLLmC, not only addresses the immediate challenges of LLM customization but also promotes scalability and operational efficiency. However, the methodology's primary limitation lies in the need for extensive testing, validation, and broader adoption across different domains.

---

[*] Corresponding author: fitsilis@uth.gr

# 1 INTRODUCTION

Software development today is facing unique challenges introduced by the fast-changing businesses environment. Among them, the rapid incorporation of Large Language Models (LLMs) brings about both revolutionary possibilities together with unique challenges in our working model. As companies progressively depend on LLMs to improve decision-making, systematize workflows, and generate innovation, the necessity to effectively manage and advance these models grows extremely important (Vaswani et al., 2022; Alammar, 2018). This is where DevOps—a set of practices designed to automate and integrate the processes between software development and IT operations—plays a critical role.

DevOps (Kim et al., 2021), traditionally focused on improving and automating the continuous delivery of software, must evolve to address the specific demands of LLMs. These models are not only complex in their structure but also require continuous data feeding, tuning, and maintenance to stay relevant and effective. The application of DevOps practices to the lifecycle management of LLMs can significantly enhance the agility and responsiveness of teams in deploying updates and improvements, thus reducing time-to-market and operational risks.

However, adapting DevOps for LLMs involves more than just a straightforward extension of existing practices. It requires a deep integration of data science workflows within the DevOps pipeline to accommodate continuous learning and model evolution. This integration must ensure that updates to LLMs, whether they are for training on new data sets or improving algorithmic efficiency, are seamlessly deployed and monitored without disrupting the operational stability of the application (Lwakatare, 2020).

Moreover, the unique characteristics of LLMs, such as their dependency on large-scale data and their sensitivity to data quality and configuration changes, introduce new challenges in version control, testing, and deployment. These challenges necessitate a rethinking of traditional DevOps tools and strategies, such as the implementation of specialized environments for training and simulation, enhanced monitoring capabilities to track model performance in real-time, and robust rollback mechanisms to mitigate potential failures (Hadi et al., 2023).

In conclusion, the customization of DevOps to suit the development and operationalization of LLMs is not merely an enhancement of existing software practices but a requisite evolution. As this paper will explore, building robust, scalable, and efficient DevOps pipelines capable of supporting the dynamic nature of LLMs is essential for businesses looking to leverage the full potential of AI technologies but as well the accuracy of the provided outputs. This adaptation not only promises significant operational efficiencies but also a competitive edge in the AI-driven market landscape.

The remaining of the paper is organized as follows. Section 2 provides a detailed background on the key concepts of DevOps, Large Language Models (LLMs) and of similar approaches indenting to integrate ML and DevOps. Further, this section also discusses the integration of ontologies and prompt engineering within the DevOps framework. Section 3 outlines the research methodology used in this study, including the review of existing DevOps practices and the development of a customized DevOps pipeline for LLMs. Section 4 presents the proposed model, named DOLLmC[1] (DevOps for Large Language Model Customization), for adapting DevOps to LLM customization, detailing each stage of the process from data acquisition to continuous evaluation and improvement. Section 5 concludes the paper by summarizing the findings, discussing the limitations, and suggesting directions for future research.

## 2  BACKGROUND

### 2.1  Key concepts of DevOps software development

DevOps is an acronym derived from the combination of software development (Dev) and the operation of the digital services (Ops). It refers to a collaborative approach whereby the software development and the service operation team, work together to provide a digital service with the maximum business value (Fitsilis, 2023; Kim et al., 2021). If one wanted to describe the three main components of the DevOps approach these would be:

- **Close Team Collaboration**: Ensuring seamless interaction between development and operations teams to enhance communication and integration.

- **Automation**: Implementing automation across all stages from coding to deployment to streamline processes and reduce manual work.

- **Continuous Monitoring and Rapid Action**: Actively monitoring system performance to detect issues early and resolve them quickly.

Further, DevOps is defined as a set of practices designed to shorten the time between introducing a change, to its implementation and the deployment of new release of the system into operation, while ensuring high quality of digital service. Therefore, DevOps is a set of best practices that integrates the two main phases of the software systems lifecycle: software development (Dev) and software system operations (Ops). As such, it is considered as a modern approach that aims to shorten the systems development lifecycle and continuously provide high-quality software (Humble & Kim, 2018).

---

[1] The acronym DOLLmC (DevOps for Large Language model Customization) is pronounced as doll-see.

Especially today, the DevOps approach is absolutely essential, since software systems are vital to the operation of all organizations and especially complex systems, they need to operate 24/7, with pressing and changing needs. The DevOps approach eliminates the gaps between the development team and the operation team providing the digital service by creating a single team.

DevOps emphasizes building trust and a better relationship between software developers and system administrators. This significantly helps the business align technical with business requirements (Freeman, 2019; Hernantes, 2016). These two teams have a different strategic approach since the software development team is usually an innovative team focused on changing and creating something new, as opposed to the digital service operation team, which is oriented to stability, uninterruptedness, repeatability, which is required to operate a service (Fitsilis, 2022).

The DevOps approach is usually depicted as a never-ending loop involving the following steps:

- **Plan**: At this stage, teams identify business needs and collect feedback from end users. They create a project roadmap in order to maximize business value and deliver the desired product.
- **Create:** This is where the code is developed. At the end of development, the system code is stored in the shared code repository.
- **Verify**: Once the software system is ready and completed, it is first tested in the test environment.
- **Packaging**: Code packaging is a key step in the DevOps software development lifecycle. It is the process of obtaining source code, finding any dependencies required, and converting it into code that can be executed on the production system.
- **Release**: Once the software has passed all test, the operation team schedules the installation of the new version the production system based on the urgent nature of the release or more generally depending on the needs of the organization.
- **Configuration management**: At this stage the software is under the control of the configuration system, while the Configurable Item (CI) has been identified, the configurations are stored in the relevant tool, the code is stored in the code repository as well as the other artifacts, etc. Information infrastructure description technologies with code such as Infrastructure-as-Code (IaC) help significantly in the automated creation of the production environment. Then the software is released with the help of related tools.
- **Monitoring**: At this stage we monitor the operation of the system by collecting data, either about customer behavior, application performance, etc. Monitoring the overall operating environment helps DevOps teams find issues where they affect both system operation and DevOps team productivity.

## 2.2 Large Language Models (LLMs)

Large language models (LLMs) represent a significant advance in AI, particularly in natural language processing. Foundation models and Generative Pre-trained Transformer models (GPTs) are currently regarded as the state-of-the-art in artificial intelligence research and applications (Kolides et al. 2022; Foster, 2022). Foundation models include a wide range of pre-trained AI models, including GPT[2], Gemini[3], Copilot[4], Jasper[5], Perplexity.ai[6], etc., designed for providing solutions in various fields. They are distinguished for their flexibility and adaptability, able to handle tasks beyond text production such as sentiment analysis, image recognition, abstract creation, plagiarism checking, literary text creation, customer service as well as creating more interactive and personalized user experiences (Hadi et al, 2023), etc. Modern LLMs use transformer architecture to improve performance (Vaswani et al., 2017; Alammar, 2018). These models are pre-trained in huge sets of text, allowing them to learn linguistic patterns and evolve in text production. They can then be configured for various tasks by providing custom training. This training enables them to produce text that is not only coherent and contextual, but also indistinguishable whether the text is written by humans or machines (Chang et al., 2023).

The application of LLMs in various fields could revolutionize the way information is used in this industry. By chatting with an AI-based chatbot powered by an LLM, users could access a wealth of knowledge tailored to their specific needs and conditions, all in real-time. LLMs have the potential to improve productivity and sustainability, but also to empower users with the knowledge and tools they need to make informed decisions about their work (Brown et al., 2020), thereby promoting innovative ideas and practices.

LLMs such as GPT-3, GPT-4, Gemini, CLIP, etc. bring transformative capabilities to text creation and comprehension, but face significant challenges (Bender, Gebru, McMillan-Major, &; Shmitchell, 2021). Bias in favor of specific recommendations in training datasets can lead to skewed recommendations, a critical issue in precision-dependent sectors, where inaccuracies can have tangible negative impacts. The nature of the "black box" complicates interpretability, making it difficult for users to trust or understand AI's advice, especially when managing production or resources (Rudin, 2019). Moreover, the significant computing resources needed to develop these models raise sustainability and accessibility concerns, especially for small-scale businesses or in areas with limited technological infrastructure. Addressing these challenges is essential for LLM models to be leveraged effectively and become a useful tool for industry professionals (Tian et al., 2024; Hadi et al, 2023).

---

[2] https://chat.openai.com/
[3] https://gemini.google.com/
[4] https://copilot.microsoft.com/
[5] https://www.jasper.ai/
[6] https://www.perplexity.ai/

## 2.3 The integration of DevOps and Machine Learning

Machine Learning (ML) and DevOps are two fundamental components in the modern software development landscape. The process of developing, deploying, and maintaining ML models involves numerous stages, including data collection, preprocessing, model training, evaluation, and deployment. MLOps, or Machine Learning Operations, is an emerging field that bridges the gap between ML model development and operations, aiming to automate and streamline the end-to-end lifecycle of ML models. It is an extension of DevOps tailored to the needs of ML projects, integrating ML systems with continuous integration/continuous delivery (CI/CD) pipelines, automated testing, and monitoring systems. MLOps brings together data scientists, ML engineers, and operations teams to improve collaboration and operationalize ML models effectively (John, Olsson, & Bosch, 2021).

MLOps can be defined as a set of practices and tools designed to deploy and maintain ML models in production reliably and efficiently. It encompasses various aspects such as CI/CD automation, workflow orchestration, reproducibility, versioning of data, model, and code, collaboration, continuous training, and monitoring (John, Olsson, & Bosch, 2021). The primary goal of MLOps is to ensure that ML models are not only developed but also deployed and maintained effectively, thus enabling continuous innovation and improvement.

Implementing MLOps offers significant advantages for managing machine learning workflows, but it's not without its challenges (Mehmood, Sabahat, & Ijaz, 2024). The inherent complexity of ML systems, encompassing various components like data pipelines, feature engineering, model training, and deployment, makes integration into a cohesive MLOps framework difficult (Cruz López, 2023). Ensuring data quality, consistency, and security is another critical hurdle. MLOps requires robust data pipelines capable of handling diverse and large datasets (Shankar et al., 2024). Scaling ML models for production environments involving large volumes of data and requests demands substantial computational resources and efficient infrastructure management (Ahsan et al., 2021). Fostering collaboration between data scientists, ML engineers, and operations teams is vital for successful MLOps implementation. This often necessitates a shift in mindset and practices to align with MLOps principles (Honkanen, Odwyer, & Salminen, 2022).

In conclusion, MLOps is a transformative practice that enhances the development, deployment, and maintenance of ML models. By integrating the principles of DevOps with the specific needs of ML systems, MLOps ensures continuous delivery of high-quality ML solutions, thereby driving innovation and operational efficiency. However, its successful implementation requires addressing several technical and organizational challenges.

## 2.4 Ontologies

Ontologies provide a structured framework for the representation and management of knowledge, categorizing concepts and their interactions into specific domains (Tsoutsa, Fitsilis, Iatrellis, 2022). This structured knowledge is vital in all sectors for effective decision-making and knowledge discovery. The integration of ontologies with major language models (LLMs) enhances the ability of these models to understand semantics and produce information related to the problem domain, basing their predictions on domain-specific knowledge (Vamsi, Kommineni, & Samuel, 2024; Giglou, D'Souza, & Auer, 2023).

For the training and successful usage of LLMs, ontologies serve as a critical bridge between generalized learning of LLMs and the specific knowledge requirements of the agricultural sector. This integration allows the creation of context-aware recommendations that are crucial for developing successful applications. Therefore, ontologies can provide the framework and information necessary for lexical models of GPTs to better understand the relationships between different entities and concepts. This is especially useful for tasks that require reasoning or a deep understanding of a topic. Also, ontologies contain logical rules and constraints that describe how concepts are related. Incorporating these rules into GPTs can enhance their reasoning abilities, allowing them to draw logical conclusions based on the structured knowledge provided by the ontology. This could lead to more complex dialogue systems that can undertake problem-solving tasks. For example, ontologies can enrich our understanding of LLMs by providing detailed information about crop diseases, leading to better diagnosis and treatment recommendations. In addition, they can help model the effects of various agricultural practices on crop yield under different environmental conditions, thus aiding in decision-making (Elvira, Procko, Ortiz Couder, &; Ochoa, 2024).

There are three integration approaches for LLMs and ontologies into a system. These are:

a) **Knowledge infusion**: Pre-training and optimization of the lexical model in datasets enriched with ontological knowledge to integrate structured information into model parameters (Lyu, 2023; Hu et al., 2023; Vanhorn, 2023).

b) **Hybrid Models**: Development of hybrid architectures that combine the productive capabilities of GPTs with separate modules that handle ontological reasoning and inference. This could include searching for an ontology-based knowledge base as part of the text creation process and

c) **Prompt Engineering**: Design prompts that explicitly leverage ontological concepts and relationships to guide GPT in creating text that reflects structured knowledge (Lo, 2023).

The integration of ontologies with major language models (LLMs) introduces several notable challenges. A major obstacle is the need to continuously update ontological frameworks to incorporate the latest findings and developments in the problem domain. The dynamic nature of the today's industry sector, driven by ongoing scientific discoveries, changing business and environmental conditions, and evolving technology, requires the knowledge embedded in these ontologies to be regularly updated. This ensures that the advice produced by LLMs remains accurate, relevant and in line with current best practices. However, maintaining an up-to-date ontology requires a systematic approach to monitoring research results and integrating new knowledge, which can require a lot of resources (Baldazzi, 2023; De Giacomo, 2018).

## 2.5 Prompt engineering

Prompt engineering is of particular importance as a key methodology for optimizing the operation of LLMs in various sectors. The logic is to generate input prompts to guide the LLM to produce outputs that are not only relevant and accurate, but also framed by the user's specific needs (Chu et al., 2020). Given the inherent generality of LLMs, prompt engineering becomes essential in fields where accuracy and specificity of information are of great importance, such as agriculture.

The effectiveness of AI-driven decision support can vary greatly depending on how well model responses align with the unique and complex conditions from each environment. For example, in agricultural sector, variables such as crop type, soil composition, climatic conditions, and local agricultural practices introduce a level of specificity that generic AI models may not adequately address and particularly without tailored prompts (Yang, 2024). Therefore, the engineering of the prompts serves as a bridge between the general capabilities of the LLMs and the detailed requirements of agricultural applications, ensuring that the guidance provided is applicable and beneficial to farmers.

Prompt engineering also facilitates the translation of structured knowledge, such as that encoded in ontologies, into actionable knowledge. By designing prompts that reflect the structured questions typically asked by an expert user, LLMs can leverage existing data to discover relationships that may not be immediately apparent without targeted prompts. Consequently, prompt engineering contributes to the decision-making process, making it a valuable tool. However, developing effective prompts requires a deep understanding of both LLMs and the sector under study. This highlights the interdisciplinary nature of engineering prompts where we need to combine knowledge from computer science, artificial intelligence, and other sciences (e.g. medicine, agriculture) to tailor LLMs to the sector's specific needs (Lo, 2023).

The strategic application of prompt engineering promises to enhance the adaptability and efficiency of LLMs by making these advanced AI tools more accessible and beneficial to the users' community. As research in this area continues to evolve, it is expected that prompt engineering will become increasingly sophisticated, further bridging the gap between the capabilities of artificial intelligence and practical business applications (Future work, 2023).

Therefore, creating effective prompts that elicit useful answers from LLMs (Lin, 2024) in niche fields such medicine, agriculture etc. requires a complex mix of domain expertise, language proficiency, and technical understanding of AI models. This challenge is magnified when the application domain is inherently complex (e.g. agriculture is characterized by a wide range of variables, including crop types, soil properties, climatic conditions, local and often invisible agricultural practices).

## 3 RESEARCH METHODOLOGY

The research methodology for this study involves the strategic adaptation of DevOps practices to meet the unique demands of customizing LLMs. This approach is built on the foundation of existing DevOps knowledge, combined with the specific requirements for LLM customization through the use of ontologies, knowledge maps, and prompt engineering.

The initial phase of the research involved a review of existing DevOps, MLOps practices. To understand how DevOps can be adapted for LLM customization, it was essential to identify the specific challenges associated with managing and customizing LLMs. LLMs, such as GPT-3.5, GEMINI, etc. are complex models that require continuous data feeding, tuning, and maintenance to remain effective and relevant.

One of the critical steps in customizing LLMs is the use of ontologies or/and knowledge maps. Ontologies provide a structured framework that defines the relationships between different concepts within a specific domain, facilitating the model's understanding and processing of information. Knowledge maps visually represent these relationships, aiding in the organization and retrieval of knowledge. To incorporate these elements into the DevOps framework, the research team studied existing methodologies for ontology development and knowledge mapping. The insights gained from this review were used to develop the DOLLmC process for integrating ontologies and knowledge maps into the DevOps pipeline, ensuring that the LLMs are equipped to handle domain-specific queries with high precision and relevance.

Prompt engineering is a crucial aspect of LLM customization, involving the creation of specific prompts that guide the models to generate accurate and contextually relevant responses. The research team explored various prompt engineering techniques and tools, such as Jupyter Notebooks, Google Colab, and the OpenAI Playground. By experimenting with these tools, the team developed a systematic approach to designing, testing, and optimizing prompts, ensuring that the LLMs could be effectively customized for specific domains.

The next phase involved designing a DevOps pipeline tailored to the customization of LLMs. This design was based on the integration of data science workflows within the traditional DevOps pipeline, enabling continuous learning and seamless deployment of updates. The pipeline was designed to include several key stages: data acquisition and curation, ontology development, LLM training and fine-tuning, prompt engineering, implementation and integration, continuous evaluation and improvement, and dissemination and user engagement.

The proposed pipeline was informed by the practical challenges identified during the literature review and case study analysis. For instance, in developing a domain-specific chatbot for agriculture, the pipeline was designed to incorporate real-time data from IoT sensors deployed on farms. This data was processed and analyzed to train and refine machine learning models, which were then integrated into a user-friendly chatbot interface.

To validate the proposed methodology, an experimental implementation was conducted focusing on the agricultural sector and on the tourist sector for developing a itinerary $CO_2$ calculator. This involved deploying the designed DevOps pipeline to develop a chatbot that could provide real-time advisory services to farmers based on data from various sources and market price trends.

## 4  THE DOLLmC PROCESS

The proposed process has been utilized in the context of LLM-based chatbot development, specifically designed to empower the agricultural and tourism sectors. Inventing and adopting a new process based on DevOps for chatbot development was necessary since LLM customization and refinement is a continuous, laborious task. This involves integrating data science workflows within the DevOps pipeline to ensure continuous learning and model evolution, seamless deployment of updates, and rigorous version control. Important challenges include managing the complexity of LLM structures, the need for ongoing data feeding, tuning, maintenance to maintain relevance and effectiveness, and handling the sensitivity of LLMs to data quality and configuration changes. These challenges necessitate rethinking traditional DevOps tools and strategies, implementing specialized environments for training and simulation, enhancing real-time performance monitoring capabilities, and establishing robust rollback mechanisms to mitigate potential failures.

Data science workflows within the DevOps pipeline integrate various stages of data handling and model management into the continuous integration and delivery processes. This integration allows for the seamless collection, processing, and analysis of data, alongside the training, tuning, and deployment of machine learning models. By embedding these data-driven tasks directly into the DevOps pipeline, teams can efficiently update and improve models in real-time, ensuring that data insights and model enhancements are rapidly operationalized and delivered without disrupting existing systems.

In our case and for the chatbot development focusing on the agricultural sector, a DevOps pipeline used data from various sources. In a production system data will be acquired real-time with the use of an IoT system where sensors will gather data on soil moisture, temperature, and crop health, which will then be processed and analyzed automatically within the pipeline. The aggregated data derived from the initial data were used to train and refine machine learning models that predict optimal planting times, water requirements, or pest control measures. This continuous loop of data collection, analysis, model updating, and deployment ensures that agricultural practices are constantly optimized based on the latest data, leading to increased crop yields and reduced resource wastage.

Similarly, integrating data science workflows into a DevOps pipeline can be exemplified by using market price data to guide planting decisions for the upcoming year. The system collects historical and current price data of various crops from market reports and agricultural databases. This data is then processed and analyzed within the pipeline to forecast future price trends using machine learning models. The predictive insights recommend which crops are likely to be most profitable to cultivate based on expected market conditions. Once the models are refined and validated, they are automatically deployed to provide real-time advisory to farmers through an agricultural management platform. This streamlined process ensures that farmers receive timely and data-driven recommendations, enabling them to make informed decisions about what crops to plant for the next planting season to maximize profitability.

The DOLLmC methodology embodies the spirit of DevOps by emphasizing continuous improvement, rapid deployment, and close collaboration between development and operational teams. Through iterative development, automated testing, and seamless integration, we aim to build a solution that not only addresses the immediate challenges faced by chatbot developers but also scales and evolves in anticipation of future advancements. This dynamic and integrated approach ensures that the developed chatbot remains a valuable and cutting-edge tool for its purposes, fostering enhanced productivity and sustainability through advanced AI capabilities. The following Figure outlines our DevOps-inspired, DOLLmC methodology for developing and deploying this innovative agricultural assistant.

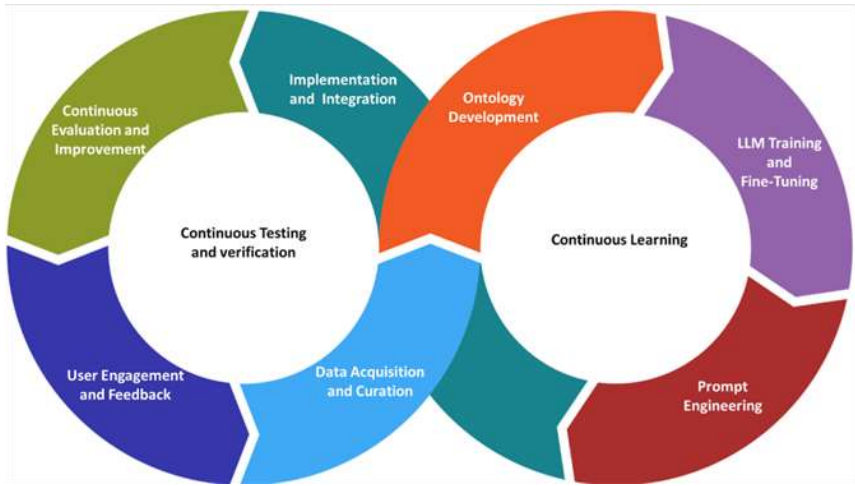

**Figure 1:** The proposed DOLLmC methodology

The breakdown of the key stages in developing and deploying a domain-specific chatbot is presented below:

1. **Data Acquisition and Curation: Building the Foundation**

- **Data acquisition** is always one of the most important steps for training and refinement of an LLM. Usually, involves:

- **Data Gathering:** The project begins by assembling a comprehensive data set that covers relevant topics within the targeted domain, such as management practices, environmental factors, and market trends.

- **Data Curation:** To ensure the chatbot delivers precise and current information, the collected data is subjected to a stringent curation process. Irrelevant or outdated information is discarded, and the data is carefully organized to minimize bias and accurately reflect the latest research and practices.

Data flows from various sources like APIs, databases, and live feeds is streamed using tools such as Apache Kafka[7] and Apache Nifi[8]. Dataflow is a software paradigm based on the idea of disconnecting computational actors by creating pipelines that can execute concurrently (also called stream processing or reactive programming). For example, Apache Kafka is an open-source event streaming platform that has been used for subscribing to streams of events and data gets published (written) to Kafka, and custom developed applications can subscribe (read) to these streams for continuous data flow. In essence, Kafka acts as a central system for creating data pipelines, facilitating the flow of information between different parts of the system. Apache NiFi complements Apache Kafka by focusing on the orchestration and management of data flows.

---

[7] https://kafka.apache.org/
[8] https://nifi.apache.org/

Obviously, different types of data require specialized handling. Structured data, are manipulated using tools like Python Pandas[9] or Apache Spark[10], which excel in processing and analyzing large datasets, and for storage, relational databases such as MySQL[11] provide a reliable solution.

On the other hand, for unstructured data NoSQL databases are used like MongoDB[12], designed to handle the flexibility and scalability needs of such data types. For data originating from IoT devices, time-series databases such as InfluxDB[13] is the best choice since they are optimized for handling time-stamped data streams. Additionally, for storing and querying knowledge graphs, graph databases like Neo4j[14] offer functionality to represent and navigate complex relationships.

2. **Ontology Development: Establishing the Knowledge Base**

During the ontology development stage of the chatbot, the team focused on constructing detailed ontologies that act as a structured semantic knowledge base, methodically categorizing and defining relationships between various domain-specific concepts and entities. This foundation enables the chatbot to process information with high precision and accuracy. Tools like Protégé[15] were employed for ontology creation, and OWL (Web Ontology Language) was used to define these ontologies in a machine-readable format. For instance, in a healthcare application, using Protégé, the team could model entities such as symptoms, diagnoses, treatments, and patient demographics, with OWL[16] for encoding these relationships. Alternatively, knowledge representation could also employ knowledge graphs or similar methodologies to encapsulate the semantic knowledge base. Additionally, LLMs can play a crucial role in assisting the creation of these ontologies by semi-automatically generating and refining them, thus streamlining the process and enhancing the chatbot's development efficiency.

3. **LLM Training and Fine-Tuning: Shaping the AI Engine**

In the LLM Training and Fine-Tuning stage, with the curated data and ontologies prepared, the project advances to the critical task of training and fine-tuning the chosen Large Language Models (LLMs). This stage is pivotal as it tailors the models to the specific domain, incorporating the structured knowledge from the ontologies to ensure the LLMs are adept at handling domain-specific queries and delivering contextually relevant and accurate responses. For model training, tools such as TensorFlow[17] or PyTorch[18] are utilized, with the option to begin with pre-trained models from Hugging Face Transformers [19] or GPT-3.5[20]. For example, in a healthcare setting, the development team might use TensorFlow to train and fine-tune a language model specifically on the healthcare ontology. This ensures the model is well-equipped to process and understand queries concerning medical conditions and treatments, starting from a pre-trained base and extending with domain-specific data to enhance precision and relevance in responses.

---

[9] https://pandas.pydata.org/
[10] https://spark.apache.org/
[11] https://www.mysql.com/
[12] https://www.mongodb.com/
[13] https://www.influxdata.com/
[14] https://neo4j.com/
[15] https://protege.stanford.edu/
[16] https://www.w3.org/OWL/

4. **Prompt Engineering: Guiding the LLMs for Optimal Performance**

In the Prompt Engineering stage, to optimize the performance of the trained Large Language Models (LLMs), the team engages in prompt engineering techniques. This involves the creation of specific prompts that mirror the typical queries and informational needs of the domain. These prompts act as guiding instructions that leverage both the training and ontological knowledge of the LLMs to deliver precise and useful responses.

For prompt engineering, a diverse array of tools and integrated environments are available, each offering unique features to enhance the workflow. Jupyter Notebook[21] provides an interactive environment ideal for real-time experimentation and visualization. Google Colab[22] extends this functionality by offering cloud-based computational resources and facilitating collaboration. The OpenAI Playground[23] allows for rapid testing and refinement of prompts through a user-friendly interface. Postman[24], an API client, excels in testing and interacting with APIs, making it particularly useful for fine-tuning prompts and integrating them into larger workflows. Additionally, specialized tools such as Promptor[25] streamline the design, testing, and optimization of prompts. These resources collectively enable efficient and effective prompt engineering, from initial experimentation to large-scale deployment.

5. **Implementation and Integration: Bridging the Gap**

---

[17] https://www.tensorflow.org/
[18] https://pytorch.org/
[19] https://huggingface.co/
[20] https://openai.com/api/
[21] https://jupyter.org/
[22] https://colab.research.google.com/
[23] https://platform.openai.com/
[24] https://www.postman.com/
[25] https://flowgpt.com/p/promptor

In the Implementation and Integration stage, the fine-tuned models are seamlessly integrated into the chatbot interface, which is designed for ease of use to allow end-users to access and interact with the chatbot via familiar digital platforms, effectively removing barriers to technical expertise. For this process, Docker[26] is used for containerization, making it easy to deploy the trained model across different environments. Kubernetes[27] plays a critical role in orchestrating these containers to scale the application as needed. Additionally, a user-friendly web interface built with a web framework like React enables end-users to interact with the chatbot on their devices without needing advanced technical skills. This combination of tools and practices ensures that the chatbot is both accessible and scalable, meeting the demands of a diverse user base.

## 6. Continuous Evaluation and Improvement

In the Continuous Evaluation and Improvement stage, a commitment to excellence is upheld through the rigorous and ongoing evaluation of the chatbot's performance after deployment. Key performance metrics such as accuracy, relevance, and user satisfaction are continuously monitored using tools like Prometheus[28] or Datadog[29], which tracks these metrics in real-time. Visualization of these metrics is handled by tools like Grafana[30], allowing the team to easily identify trends and potential issues. Additionally, open source tools like ELK Stack[31] is utilized for log management, collecting and analyzing logs to provide deep insights into operational issues. This comprehensive monitoring framework supports iterative improvements to the chatbot, which may involve updates to ontologies, refinements in LLM training, and adjustments to prompt engineering techniques. Feedback from end-users and domain experts plays a crucial role in this process, ensuring that each iteration of the chatbot more effectively meets the needs of its users and reflects the latest in domain expertise.

## 7. Dissemination and User Engagement

In the Dissemination and User Engagement stage, the focus shifts to actively promoting the chatbot within the target community through a series of workshops, demonstrations, and targeted outreach programs. These activities are essential for engaging users and ensuring that the chatbot reaches its intended audience effectively. Tools like Google Analytics[32], and HubSpot[33] play an important roles in this process. Google Analytics is used to track user interactions with the chatbot, gathering insights on usage patterns and areas for improvement. HubSpot aids in automating outreach emails and social media posts to maximize reach and engagement.

---

[26] https://www.docker.com/
[27] https://kubernetes.io/
[28] https://prometheus.io/
[29] https://www.datadoghq.com/
[30] https://grafana.com/
[31] ELK is an acronym for a family of open source tools like Elasticsearch, Logstash, and Kibana. Elasticsearch is the engine of the Elastic Stack, which provides analytics and search functionalities. https://www.elastic.co/elastic-stack

By utilizing these tools, the team actively promotes the project's value proposition, aiming to foster widespread adoption and enhance user engagement, ultimately contributing to increased productivity and sustainability within the domain. This framework, bolstered by strategic tool use and user feedback, can be adapted by software development teams to create robust, domain-specific chatbots that are agile, responsive, and tailored to meet the specific needs of any professional field or industry, ensuring a robust, efficient, and user-friendly chatbot tailored to their specific domain.

In Table 1 we summarize the key technology stack that has been used in the developed customized LLMs and it could be possibly used in similar cases.

**Table 1:** Technology stack used in each development step.

| Process Step | Tools |
|---|---|
| **Data Acquisition and Curation** | <ul><li>Data collection methods (domain-specific)</li><li>Data flow management (e.g. Apache NiFi)</li><li>Structured data manipulation (e.g. Python Pandas/ Apache Spark)</li><li>Relational databases for storing structured data (e.g. MySQL)</li><li>NoSQL databases for unstructured data (e.g. MongoDB)</li><li>Time-series databases for IoT devices data streams (e.g. InfluxDB)</li><li>Graph database for storing knowledge graphs. (e.g. neo4j)</li></ul> |
| **Ontology Development** | <ul><li>Existing domain ontologies. For example, for agricultural sector that we have studied agrovoc multilingual thesaurus34 or agronomy ontology35.</li></ul> |

---

[32] https://analytics.google.com/analytics/academy/
[33] https://www.hubspot.com/
[34] https://agrovoc.fao.org/browse/agrovoc/en/
[35] https://bigdata.cgiar.org/resources/agronomy-ontology/

|  |  |
|---|---|
|  | - Ontology development tools (e.g. Protégé)<br>- Web Ontology Languages (e.g. OWL) |
| **LLM Training and Fine-Tuning** | - Open-source LLMs (e.g., GPT-3.5, Hugging Face Transformers)<br>- Deep learning frameworks (e.g. TensorFlow, PyTorch) |
| **Prompt Engineering** | - Custom script development<br>- Jupyter Notebooks<br>- Online prompt environments (e.g. Google Colab, OpenAI Playground, Postman, Promptor) |
| **Implementation and Integration** | - Containerization tools (e.g., Docker)<br>- Cloud platforms (e.g., GCP, AWS)<br>- Development frameworks (e.g. React, Flutter) |
| **Continuous Evaluation and Improvement** | - Monitoring and analytics tools (e.g. Prometheus)<br>- Visualization tools (e.g. Grafana)<br>- User feedback mechanisms (surveys, A/B testing, in-app feedback) |
| **Dissemination and User Engagement** | - User training and onboarding materials<br>- Marketing and outreach (social media, partnerships, industry publications) |

# 5 CONCLUSIONS, LIMITATIONS AND FUTURE WORK

In conclusion, this research underscores the critical role of adapting DevOps practices to the unique requirements needed for customizing LLMs. By integrating ontologies/knowledge maps, and prompt engineering into the DevOps pipeline, we have proposed a comprehensive framework that enhances the efficiency and effectiveness of LLM deployment. The application of DOLLmC methodology in the development of domain-specific chatbots for agriculture and tourism demonstrates its potential to deliver real-time, data-driven insights and recommendations, thereby improving decision-making and operational efficiency in these sectors.

However, the research also acknowledges several limitations. The primary limitation is the need for extensive validation of the DOLLmC methodology. While our pilot/experimental implementation has shown promising results, it is crucial to conduct broader trials across different domains and environments to fully assess the scalability and robustness of the approach. Additionally, the integration of ontologies and the continuous updating of these knowledge frameworks pose significant challenges that require ongoing attention and resources.

Future work will focus on addressing these limitations by conducting extensive field tests and validation studies to refine the methodology further. It will also involve developing automated tools and processes for updating ontologies and integrating new knowledge, thereby ensuring that LLMs remain relevant and accurate over time. Moreover, future research will explore the potential of hybrid models that combine LLMs with specialized modules for ontological reasoning and inference, enhancing the models' ability to handle complex, domain-specific queries. By continuing to evolve and improve the DevOps framework for LLM customization, we aim to establish a robust, scalable, and efficient approach that can be widely adopted across various industries.